\documentclass[aps,twocolumn,superscriptaddress,prl,showpacs]{revtex4}

\usepackage{graphicx}
\usepackage{bm}

\begin{document}
\title{Anomalous Hanle Effect in  Quantum Dots : Evidence for Strong Dynamical Nuclear Polarization in Transverse Magnetic Field}



\author{O.~Krebs}
\affiliation{CNRS-Laboratoire de Photonique et de Nanostructures, Route de Nozay, 91460
Marcoussis, France}
\author{P.~Maletinsky}
\affiliation{Institute of Quantum Electronics, ETH-H\"{o}nggerberg, CH-8093, Z\"{u}rich, Switzerland}
\author{T. Amand}
\author{B. Urbaszek}
\affiliation{Universit\'{e} de Toulouse, INSA-CNRS-UPS, LPCNO, 135 Av. Rangueil, 31077 Toulouse, France}
\author{A.~Lema\^{i}tre}
\author{P. Voisin}
\affiliation{CNRS-Laboratoire de Photonique et de Nanostructures, Route de Nozay, 91460
Marcoussis, France}
\author{X. Marie}
\affiliation{Universit\'{e} de Toulouse, INSA-CNRS-UPS, LPCNO, 135 Av. Rangueil, 31077 Toulouse, France}
\author{A.~Imamoglu}
\affiliation{Institute of Quantum Electronics, ETH-H\"{o}nggerberg, CH-8093, Z\"{u}rich, Switzerland}

\def\xp{$X^{+}$ }
\def\xm{$X^{-}$ }
\def\x0{$X^{0}$ }


\date{\today}

\begin{abstract}
Hanle effect is ubiquitous in the study of spin-related phenomena and has been used to determine spin lifetime, precession and transport in semiconductors. Here, we report an experimental observation of anomalous Hanle effect in individual self-assembled InAs/GaAs quantum dots where we find that a sizeable photo-created electron spin polarization can be maintained in transverse fields as high as 1~T until it abruptly collapses. The striking broadening of the Hanle curve by a factor of $\sim 20$ and its bistability upon reversal of the magnetic sweep direction points to a novel dynamical nuclear spin polarization mechanism where the effective nuclear magnetic field compensates the transverse applied field. This interpretation is further supported by the measurement of actual electron Zeeman splitting which exhibits an abrupt increase at the Hanle curve collapse. Strong inhomogeneous quadrupolar interactions typical for strained quantum dots are likely to play a key role in polarizing nuclear spins perpendicular to the optically injected spin orientation.
\end{abstract}

\maketitle

The spin of a conduction electron confined in a semiconductor quantum dot (QD) is a good candidate  for the realization of a quantum bit in the condensed matter, mostly because of its long coherence time\,\cite{Greilich-Science07,Sci309-Petta,Koppens-Nature442,Wu-PRL07}. In a III-V semiconductor matrix however, it experiences an effective internal magnetic field via its hyperfine interaction with a finite number of $N\sim$10$^4$-10$^6$ nuclear spins, which can considerably affect its dynamics. For an unprepared nuclear spin system, statistical thermal fluctuations yield a random nuclear field with an \textit{r.m.s.} strength $\propto 1/\sqrt{N}$, amounting to a few 10 mT. This gives rise to a spin dephasing process on a ns timescale\,\cite{PRL94-Braun,PRB65-Merkulov2002}. It has recently been shown that this detrimental effect can be suppressed by exploiting the nonlinear feedback between electron and nuclear spins, which can lead to a reduction of the nuclear spin fluctuations\,\cite{Latta-NatPhy09,Vink-NatPhy09}. Similarly, creation of an effective nuclear magnetic field (Overhauser field) through dynamical nuclear polarization (DNP)\,\cite{PRL96-Lai,Krebs-CRP08} is also predicted to suppress nuclear spin mediated electron spin dephasing. In this case, the resulting finite nuclear field $\bm{B}_n$  acts back on the electron spin dynamics so that the electron-nuclei spin system exhibits a strongly non-linear dependence on external parameters. For example, when the magnetic field is applied parallel but opposite to the photo-created nuclear field a  pronounced bistability regime develops~\cite{Braun-PRB74,Maletinsky-PRB75,Tartakovskii-PRL98,Kaji-APL07,Krebs-CRP08}.\\
\indent   In a magnetic field perpendicular to the optical axis $z$, DNP is in contrast expected to vanish because of the Larmor precession of the nuclei about the transverse field, when it exceeds the electron-generated Knight field of at most a few 10~mT~\cite{Optical-Orientation,PRL96-Lai}. The subsequent depolarization of the electron spin by the external field, the so-called Hanle effect, is classically expected to follow a Lorentzian curve with a half-width given by $B_{1/2}=\hbar/(|g_e|\mu_B\tau_s^\star)$\,\cite{Optical-Orientation} where $g_e$ is the Land\'{e} g-factor of the QD electron, $\tau_s^\star$ the electron spin lifetime and $\mu_B$ the Bohr magneton. Hanle effect measurements on individual GaAs QDs   have  confirmed the absence of nuclear effects under 40~kHz polarization modulation of the exciting light\,\cite{PRL94-Bracker}. 
However, recent experimental studies on ensembles of self-assembled InP/InGaP QDs\,\,\cite{Korenev-PRL07} have revealed a significant broadening of the expected Hanle curve when the excitation polarization is constant. This  behavior has been attributed to the existence of a strong longitudinal nuclear magnetic field, $B_{n,z}$, developing under circularly polarized excitation, even in the presence of a transverse field, thanks to the strain-induced quadrupolar splittings (QS) of the nuclear spin states.\\
\indent In this Letter we investigate the Hanle effect of electrons in individual, singly-charged InAs/GaAs QDs. We find drastic distortions of the corresponding depolarization curves characterized by a  $\sim$20 times broadening, an abrupt drop at a critical field and an hysteresis upon sweeping back the field. Our analysis shows that this anomalous Hanle effect  results from  a nuclear effective magnetic field compensating almost totally the applied transverse field up to $\sim$1~T. In contrast to the conclusion of Ref.~\cite{Korenev-PRL07}, we infer that only a small longitudinal  nuclear field subsists. Yet, the role of nuclear QS  seems still essential to understand  the  conversion of the longitudinal electron spin polarization into a transverse nuclear field.\\
\begin{figure}[t]
\includegraphics[width=0.48 \textwidth,angle=0]{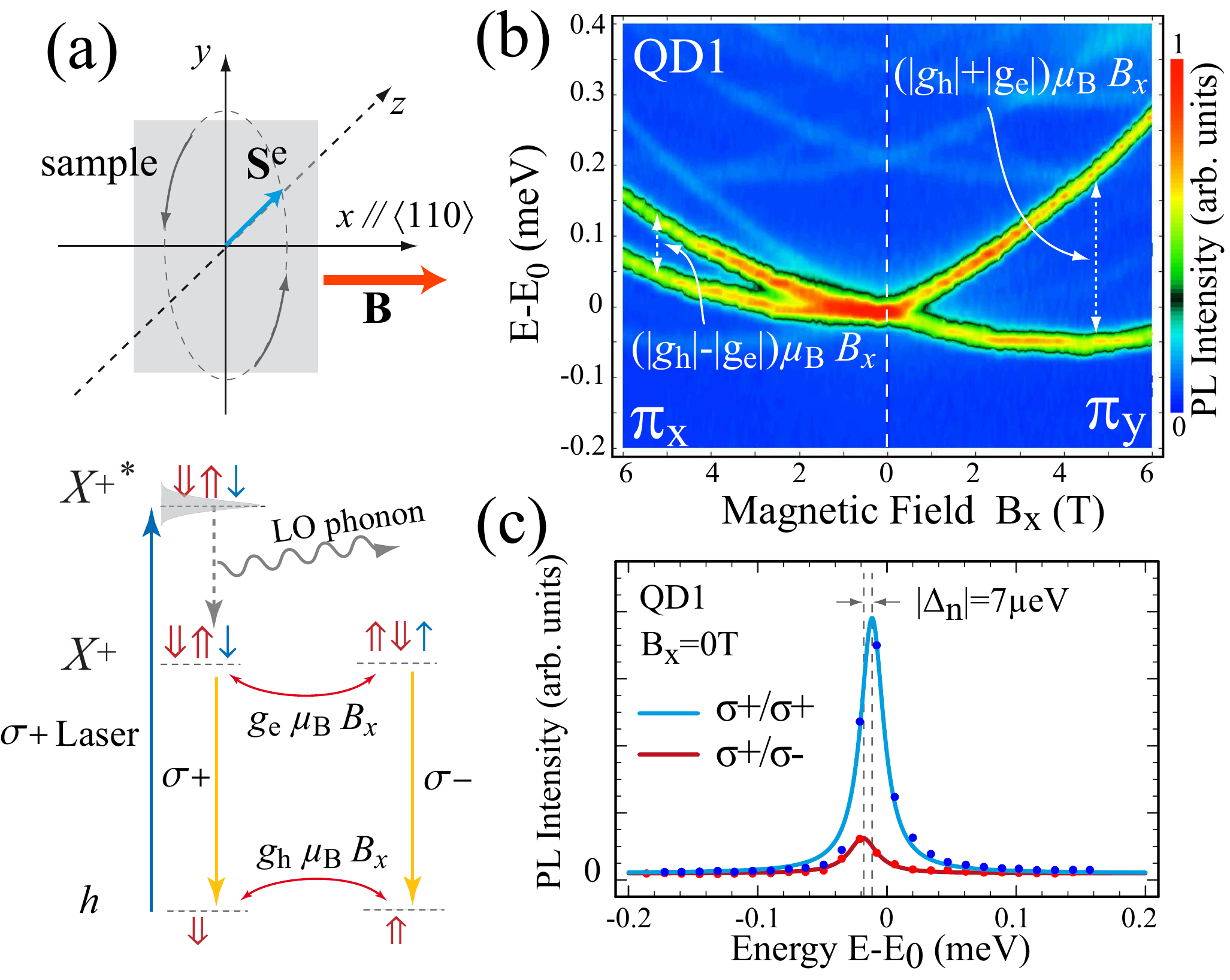}
\caption{(Color online) (a) Experimental geometry for Hanle effect measurements and schematics of the $X^+$ optical orientation for electron spin initialization. (b) Dispersion of the PL intensity around $E_0=$1.3522~eV for QD1 in a transverse magnetic field under linearly-polarized excitation, measured in a linear polarization basis as indicated. $|g_e|$ and $|g_h|$ can be extracted from the Zeeman splitting of the $X^+$-line. (c) Polarization resolved $X^+$ PL line at $\bm{B}_{\rm ext}=0$ under constant $\sigma^+$ excitation at 1.396~eV.} \label{fig1}
\end{figure}
\indent We have investigated two samples (A and B) grown by molecular beam epitaxy on a semi-insulating GaAs [001] substrate and consisting of a single layer of self-assembled InAs/GaAs QDs. In sample A, the QDs are positively charged with one excess hole due to residual doping, while sample B  has a diode structure enabling us to control the charge state with a gate voltage\,\cite{Warburton-Nat00}. The $\mu$-photoluminescence (PL) spectroscopy of individual QDs was carried out in split-coil magneto-optics cryostats at T=1.8~K with optical setups  providing a typical spectral resolution of 25~$\mu$eV.  In contrast to ensemble measurements, our experiments allow thus for a direct determination of nuclear fields through the corresponding excitonic lineshifts.\\
\indent To measure the Hanle effect of electrons ($e$), individual QDs charged with a single excess hole ($h$) are optically excited with a quasi-resonant circularly-polarized light propagating along the $z$ direction. This creates positively charged excitons $X^+$ (``trions'') consisting of 1 electron and 2 holes. Owing to optical selection rules and to the spin conservation by phonon-assisted relaxation, the spin of the photo-created electron  can be prepared in state $\uparrow$ or $\downarrow$, while both holes are paired in a spin singlet state, as schematically illustrated in Fig.~\ref{fig1}(a). 
The optical orientation of the $e$-spin, $\langle S_z^e\rangle$, can be  monitored by the PL circular polarization $\mathcal{P}_c$  through the relation  $\mathcal{P}_c =-2\langle S_z^e \rangle$~\cite{Krebs-CRP08,PRL96-Lai}.
In zero field, $\mathcal{P}_c$ achieved under quasi-resonant excitation reaches more than 70\%, as shown in Fig.\ref{fig1}(c). Since in InAs QDs the electron spin lifetime $\tau_s^\star$ is essentially determined by the trion radiative lifetime $\tau_r\approx1$~ns, the half-width of the Hanle curve, $B_{1/2}$, is expected to be $\sim$30~mT by taking a typical value of $|g_e|\sim0.5$. The exact value of $g_e$ can be determined by fitting the characteristic $X^+$ Zeeman splitting into 4 lines,  which are resolvable in strong transverse magnetic fields (Fig.~\ref{fig1}(b)).
The inner ($\pi_x$-polarized) and outer ($\pi_y$-polarized) lines which emerge from the zero-field $X^+$ line are split by the sum and difference of the $e$ and $h$ Zeeman splittings (see Fig.~\ref{fig4}(a)). In accordance with previously reported $|g_e|$ values\,\cite{Cond-mat_Krebs}, we extract  $|g_e|$=0.34 and $|g_h|$=0.6 from this measurement. We note that the identical analysis on the QD studied in sample B yields $|g_e|$=0.46 and $|g_h|\approx0$~\cite{SampleBgh}. The Overhauser shift $\Delta_n$ at $\bm{B}_{\rm ext}=0$ (Fig.~\ref{fig1}(c)) corresponds to a nuclear magnetic field $B_{n,z}=\Delta_n/(|g_e|\mu_B)=$0.35~T experienced by the electron. We would therefore expect possible anomalies in the Hanle curve for a range of transverse magnetic fields of at most a few $100~$mT.\\
\begin{figure}[t]
\includegraphics[width=0.4 \textwidth,angle=0]{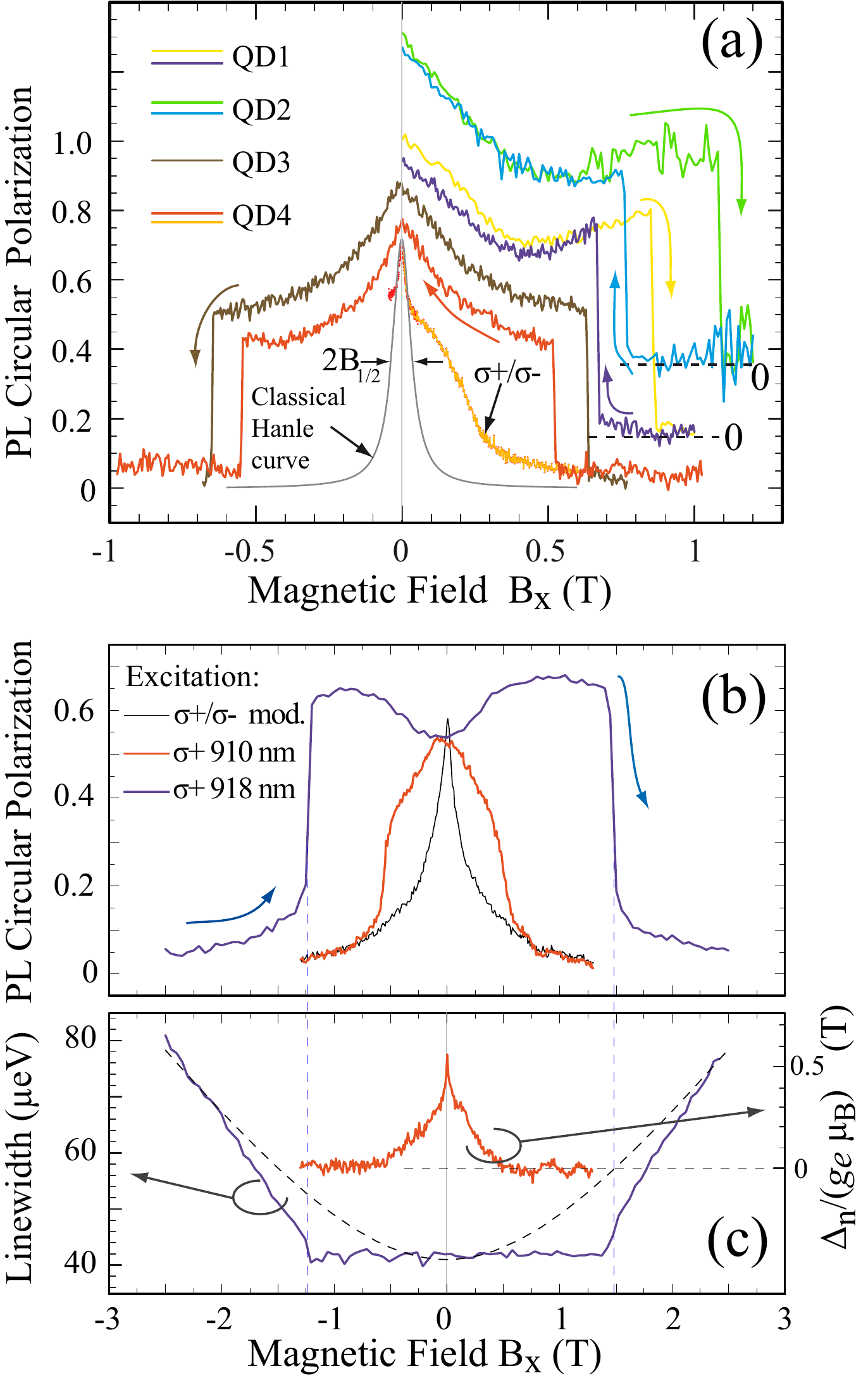}
\caption{(Color online) (a) Hanle depolarization curves of $X^+$ trions in 4 different QDs in sample A,  under a constant $\sigma^+$ polarized excitation. QD4 was also measured under  $\sigma^+/\sigma^-$-modulated excitation. The field was swept as indicated by arrows.  Curves for QD1 and QD2 have been vertically shifted for clarity. The  classical Hanle curve is that expected in absence of nuclear effects.  (b) Hanle depolarization curves of an $X^+$ trion in sample B under different excitation conditions as indicated. (c) $X^+$ apparent linewidth and   nuclear field component $B_{n,z}$ estimated from the $\sigma^+/\sigma^-$ splitting $|\Delta_n|$. The dashed line is the evolution of effective linewidth  due to $e$ Zeeman splitting.} \label{fig2}
\end{figure}
\indent Figure \ref{fig2} shows typical Hanle depolarization curves for samples A and B. The discrepancy to the expected Lorentzian curve is considerable. The width of Hanle curves is increased by more than one order of magnitude with respect to the expected $B_{1/2}$ and their shape is clearly distorted from a Lorentzian. In sample A (Fig.~\ref{fig2}(a)), $\mathcal{P}_c$ decreases slowly (over ~$\sim$0.5~T) to about half of its initial value, and then remains essentially constant until it abruptly collapses to zero. This drastic non-linearity indicates a likely bistability of the electron-nuclear spin-system which is indeed revealed by a hysteretic behavior when sweeping $B_x$ back to zero (see QD1 and QD2). This  behavior differs noticeably from the DNP non-linearity  observed  in a longitudinal magnetic field which exhibits a strong asymmetry with respect to the field direction~\cite{Braun-PRB74,Maletinsky-PRB75,Tartakovskii-PRL98,Kaji-APL07,Krebs-CRP08}.  Here, the non-linearity is found for both directions of the transverse field (see QD3 and QD4). Sample B exhibits a very similar dependance of $\mathcal{P}_c$ on the magnetic field (Fig.\,\ref{fig2}(b), red curve). In addition, we find that the qualitative behavior of the Hanle curve depends sensitively on the distinct resonances used for QD excitation as shown in Fig.\,\ref{fig2}(b). We assign these qualitative changes to the various QD excitation channels that are involved under quasi-resonant excitation~\cite{eRelax}.\\
\indent The anomalous Hanle curves and the magnetic field range over which these anomalies occur, suggest that the effective magnetic field experienced by the QD electron is strongly influenced by nuclear fields. This conclusion is further supported by measuring Hanle depolarization curves where the excitation polarization is modulated between $\sigma^+$ and $\sigma^-$ at a few kHz - significantly faster than the timescale of DNP buildup~\cite{Maletinsky-PRL07} (Fig.~\ref{fig2}(a)-(b)). As a result, $\langle S_z^e \rangle$ averages to zero and nuclear fields should be strongly suppressed. We find that in this case, the strong singularities of anomalous Hanle curves indeed vanish and that $B_{1/2}$ is reduced to $\sim$0.2~T. Even under modulated excitation helicity, however, the normal Hanle lineshape is still not recovered. It is known that under such excitation nuclear spin effects can not be completely neglected. Processes like ``resonant spin cooling''\,\cite{Optical-Orientation} could still affect the shape of the Hanle curve - a detailed description of these effects however is out of the scope of this paper.\\
\indent  The fact that under constant excitation polarization, $\mathcal{P}_c$ is substantially preserved  requires the total magnetic field $\bm{B}_T=\bm{B}_n+\bm{B}_{\rm ext}$ to have a dominant  $z$-component ($|B_{T,z}|\geq |B_{T,\perp}|$) or a small in-plane strength ($|B_{T,\perp}|\leq B_{1/2}$). This can be achieved either by (a): a strong nuclear field along the $z$-axis, such that $|B_{n,z}|\geq |B_x|$, or (b): by a nuclear field nearly anti-parallel to $B_x$ such that $|B_{n,x}+B_{x}|\leq \rm{max}(|B_{n,z}|,B_{1/2})$. Option (a) clearly holds in zero applied field; in the presence of a transverse external magnetic field, however, one would expect $B_{n,z}$ to vanish due to Larmor precession. As discussed  in Ref.~\,\cite{Korenev-PRL07}, quadrupolar interactions of nuclear spins in QDs could ``stabilize'' the nuclear spins and allow for a finite $B_{n,z}$, even in the presence of $B_{x}$. As long as $|B_{n,z}|\geq|B_x|$, the Hanle effect should therefore be suppressed giving rise to a broadening of the depolarization curve of the order of $|B_{n,z}|$. However, the initial strength of the nuclear field ($\sim 0.5$~T) seems rather insufficient to explain the observed broadening up to $\sim$1~T. 
Furthermore, the evolution of the  $\sigma^+$-$\sigma^-$ splitting $\Delta_n$ (Fig.~\ref{fig2}(c)) indicates a dramatic decrease of $|B_{n,z}|$ with the applied field, which definitely rules out option (a).  
 We therefore conclude that the observed anomalous Hanle effect is a result of an in-plane nuclear magnetic field which develops antiparallel to the applied field. While the detailed microscopic process for the establishment of this nuclear spin polarization is still unclear, several observations qualitatively support our scenario: (i) in the presence of a tilted magnetic field $\bm{B}_T$, $\bm{S}^e$ can acquire a finite average value along $x$ (see Fig.~\ref{fig4}(c)), which can be subsequently transferred to the nuclear spins via DNP, (ii) the nuclear spin component $B_{n,x}$ is conserved by the applied field; besides, in-plane strain in the QD lattice and alloy disorder due to Ga and In intermixing could lead to a further stabilization of the nuclear field along $x$\,\cite{Korenev-PRL07,PRB71-Deng}, and (iii) compensating the external field with a nuclear field reduces the total $e$ spin splitting and therefore favors a high DNP rate\,\cite{PRB-Eble,Maletinsky-PRB75}. Scenario (b) is further  supported by a measurement of the $X^+$-linewidth in $\sigma^\pm$ polarization as shown in Fig.~\ref{fig2}(c). It remains constant (limited  by  spectral resolution) until the critical field of polarization collapse is reached. At this point the apparent linewidth undergoes an abrupt increase due to the restoration of the trion Zeeman splitting when the  in-plane nuclear field vanishes.\\
\indent To really demonstrate the existence of an in-plane nuclear field, we measured  QD1  in the basis of linear polarizations ($\pi_x$ and $\pi_y$) while keeping a circularly-polarized  excitation. This choice corresponds to the proper selection rules of $X^+$ transitions for $B_{n,x}$=0, as illustrated  in Fig.~\ref{fig4}(a), and thus improves the resolution of their splittings. When the trion circular polarization vanishes  at 0.85~T, we observe a clear discontinuity of the $\pi_x$- and $\pi_y$-polarized splittings  as deduced from a double Lorentzian fit of the experimental trion line, see Fig.~\ref{fig4}(b). The $\pi_y$ splitting undergoes an increase by 14~$\mu$eV, while the $\pi_x$ splitting is reduced by at least 10~$\mu$eV to $\approx20~\mu$eV (the validity limit of our fit). These jumps reflect the situation depicted in Fig.~\ref{fig4}(a). Below 0.85~T, the nuclear field almost exactly compensates the applied field such that the trion splitting originates mostly from the $h$ spin splitting $|g_h|\mu_B B_x$. Above 0.85~T, the nuclear field has essentially vanished so that the $\pi_x$ ($\pi_y$) splitting is decreased (increased) by $|g_e| \mu_B B_x$. As shown in Fig.~\ref{fig4}(d) this interpretation agrees well with the $g$-factors determined above in a strong field (dashed lines). When the magnetic field is swept back to zero, we observe the opposite jumps yet shifted to a lower field (0.65~T) in agreement with the polarization hysteresis of QD1. As a control experiment, we also checked that under linearly-polarized excitation (i.e. no nuclear field) the $\pi_y$ splitting evolves linearly as $(|g_e|+ |g_h|)\mu_B B_x$.\\
\indent Finally, the likely scenario for DNP in a transverse magnetic field is depicted in Fig.~\ref{fig4}(c). Assuming a fast $e$ Larmor precession (i.e. $|\bm{B}_T|> B_{1/2}$), the average $e$ spin $\langle \bm{S}^e\rangle$ would be roughly aligned  along the total field axis. Thereby, the  nuclear field  optically generated through DNP is antiparallel to $\langle \bm{S}^e\rangle$, because of the negative sign of $g_e$. Under the action of $\bm{B}_x$ its $z$-component should vanish, but thanks  to  the  nuclear QS  in   biaxially  strained InAs QD~\cite{Korenev-PRL07,PRB71-Deng}, a  finite  $B_{n,z}$ component can  subsist. The resulting average nuclear field $\langle \bm{B}_n\rangle$ is thus  neither co-linear to $\langle \bm{S}^e\rangle$ nor to the applied field. The  $B_{n,z}$ component is essential in this scenario to maintain the out-of-plane component of $\bm{B}_T$. It allows for spin transfer from $\langle S^e_z\rangle $ to $\langle S^e_x\rangle $ giving rise to the in-plane nuclear field $B_{n,x}$  which countervails the applied field. 
\begin{figure}[t]
\includegraphics[width=0.4 \textwidth,angle=0]{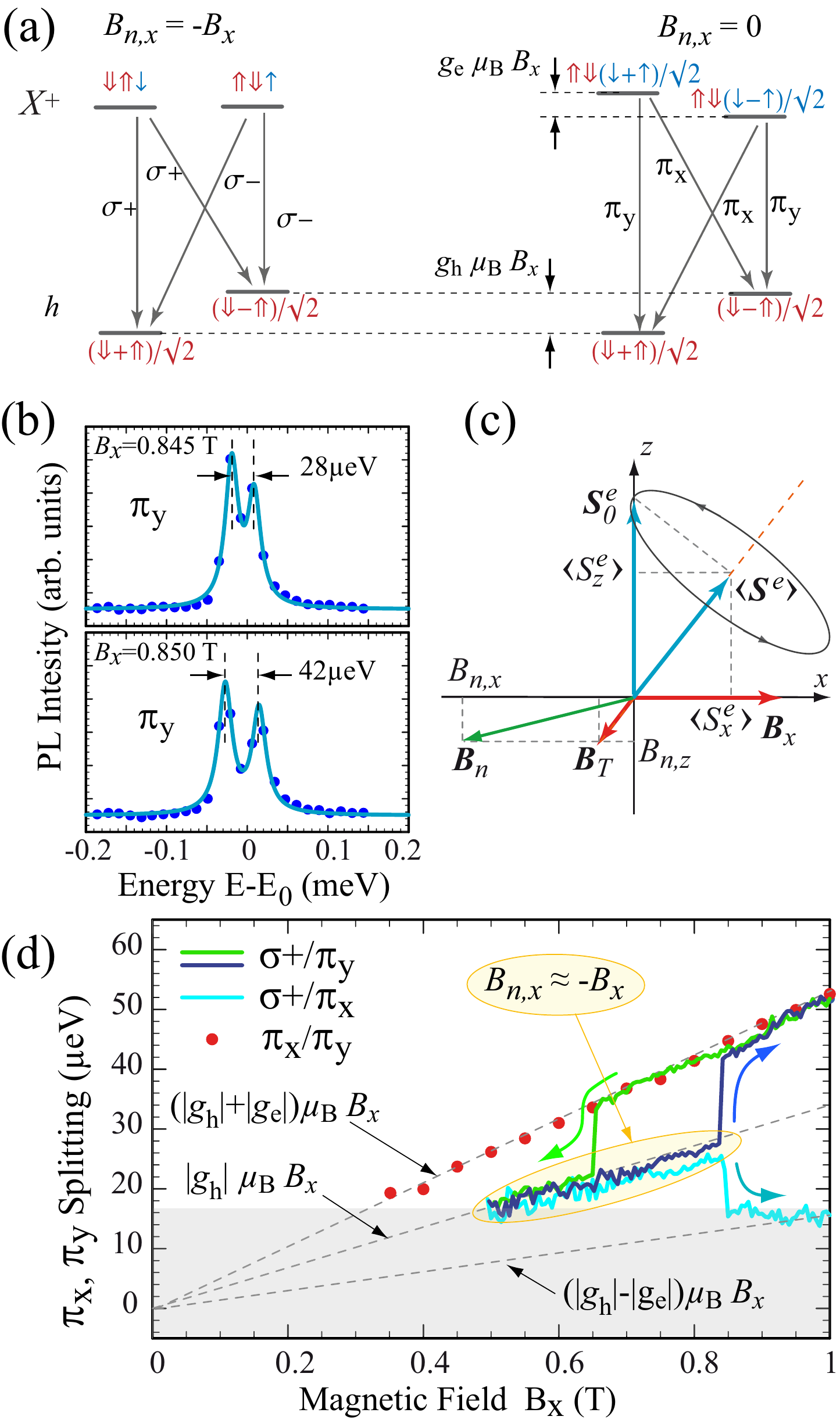}
\caption{(Color online) (a) Schematics of  $X^+$ optical transitions in a transverse magnetic field $B_x$ for both limit cases  $B_{n,x}=-B_x$  and $B_{n,x}=0$, assuming $B_{n,z}=0$. (b) Fit of the QD1 $X^+$ PL line measured in  linear $\pi_y$ polarization by a  Lorentzian doublet on both sides of the circular polarization collapse. (c) Diagram  of $e$ spin and  nuclear   field components  leading to the experimental observations. (d) Splitting of the QD1 $X^+$   line measured in $\pi_x$ or $\pi_y$ polarization above  the confidence limit of the fit (gray-shaded area). The dashed lines represent the $h$ and $e\pm h$ splittings as indicated.} \label{fig4}
\end{figure}\\
\indent In conclusion, we have observed a spectacular broadening and hysteretic behavior of Hanle depolarization curves for  electrons in InAs quantum dots. The analysis of these data evidences a novel mechanism of dynamical nuclear polarization characterized by a strong nuclear field almost perpendicular to the optically pumped electron spin orientation and antiparallel to the applied field. We  suspect the strong nuclear quadrupolar shifts arising from the quantum dot biaxial strain to play a crucial role in  the establishment of this nuclear field.  Further theoretical and experimental investigations are  necessary to understand how this counterintuitive  regime of the coupled electron-nuclear spin system develops in InAs quantum dots. This could reveal interesting new effects such as self-sustained oscillations of the nuclear spin-polarization\,\cite{Optical-Orientation}.\\

\begin{acknowledgments}
This work has been supported by ANR contract MOMES, the r\'{e}gion Ile-de-France and the NCCR nanoscience. P.~M. and A.~I. acknowledge A.~Badolato for sample growth.
\end{acknowledgments}


\bibliographystyle{apsrev}

\begin{thebibliography}{24}
\bibitem[{Sam()}]{SampleBgh}
\bibinfo{note}{From the Hanle-curve for negative charging, we deduced an upper
  bound of $|g_h|<0.03$ . The variability of $g_h$ between QDs results from the
  actual heavy-/light-hole mixing.}
\bibitem[{eRe()}]{eRelax}
\bibinfo{note}{As the orbital wavefunctions of the QD excited states have a
  larger lateral extent than in the ground state, electrons in these levels
  likely experience a reduced nuclear field and could therefore be partially
  depolarized by the transverse field during relaxation.}
\end{thebibliography}

\end{document}